\documentclass{article}
\usepackage{spconf,amsmath,graphicx,amsfonts, hyperref}
\usepackage{xcolor}
\usepackage[utf8]{inputenc}
\usepackage{array}
\usepackage{makecell}
\usepackage{booktabs}
\usepackage{multirow}
\newcommand{\ra}[1]{\renewcommand{\arraystretch}{#1}} 

\title{Multimodal Estimation of Change Points of Physiological \\Arousal in Drivers}

\name{Kleanthis Avramidis\qquad Tiantian Feng\qquad Digbalay Bose\qquad Shrikanth Narayanan}
\address{Signal Analysis and Interpretation Lab, University of Southern California, Los Angeles, CA 90089}

\begin{document}
\ninept
\maketitle

\begin{abstract}
Detecting unsafe driving states, such as stress, drowsiness, and fatigue, is an important component of ensuring driving safety and an essential prerequisite for automatic intervention systems in vehicles. These concerning conditions are primarily connected to the driver's low or high arousal levels. In this study, we describe a framework for processing multimodal physiological time-series from wearable sensors during driving and locating points of prominent change in drivers' physiological arousal state. These points of change could potentially indicate events that require just-in-time intervention. We apply time-series segmentation on heart rate and breathing rate measurements and quantify their robustness in capturing change points in electrodermal activity, treated as a reference index for arousal, as well as on self-reported stress ratings, using three public datasets. Our experiments demonstrate that physiological measures are veritable indicators of change points of arousal and perform robustly across an extensive ablation study.
\footnote{Code and results available at https://github.com/usc-sail/ggs\_driving}
\end{abstract}


\section{Introduction}
\label{sec:intro}

Driving safety has become a popular field of research and a critical investment in the automobile industry. Yet, and despite the technological advances, millions of car accidents still occur every year, with thousands of casualties in total, as reported by the World Health Organization \cite{who2018safety}. Among the most common causes of car accidents is unsurprisingly due to the driver's behavior. Driving under stress or fatigue, caused either by the workload of driving itself or by external factors that affect a driver's life, is a determining factor in these incidents. The affected drivers are more prone to risky behavior, such as getting easily distracted or expressing road rage \cite{matthews1998personality, zhang2019driving}, rendering them a danger both to themselves and the nearby drivers. Even when it may not lead to accidents, stress or fatigue could still pose an impact on drivers' mental health. Research has linked driving stress to risk of depression, and decline of satisfaction \cite{gee2004traffic}. Fatigued drivers also report degraded sleep quality and general well-being \cite{useche2018work}.

Stress can be described as a psycho-physiological reaction to a variety of factors that attempt to interfere and deregulate human experience in everyday life. This reaction is intrinsically connected to the activity of the autonomous nervous system, which can be tracked by modern physiological sensors. Stress has been connected to sweat gland activity \cite{stelmack2004psychobiology} that is measured by electrodermal activity (EDA) sensors. Heart activity, breathing rate, and blood pressure have also been found to correlate with stress events \cite{nvemcova2020multimodal}. Given the wide availability of sensors to monitor these physiological signals, many signal processing systems have been proposed \cite{sarker2016finding,fogarty2004examining} for stress detection, with the overarching goal of providing just-in-time intervention. One of the major challenges in this process is to determine the appropriate moments for the system to trigger just-in-time intervention by effectively detecting a prominent change in the driver's stress or arousal level. Detecting dynamic changes in stress while driving is an under-explored topic in the literature.

\begin{figure*}[h]
    \centering
    \includegraphics[width=0.98\linewidth]{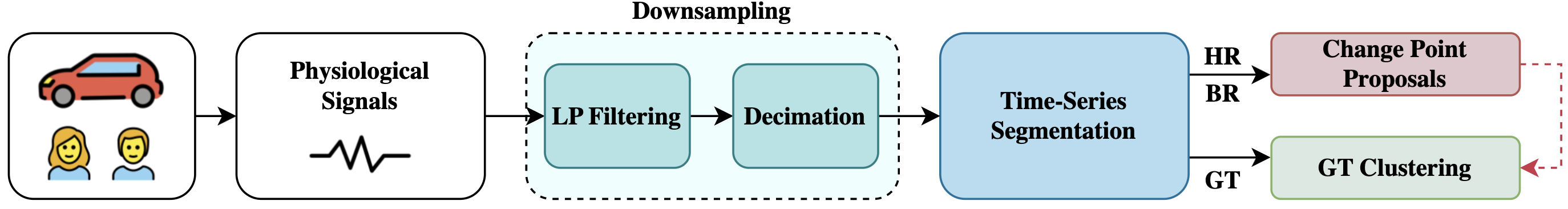}
    \vspace{-2mm}
    \caption{The proposed framework for multimodal change point detection in the driving setting. Physiological signals are downsampled and timestamps of change points are provided by GGS. HR stands for Heart Rate, BR for Breathing Rate, and GT for the Ground Truth.}
    \label{fig:change_point_framework}
    \vspace{-3mm}
\end{figure*}

In this study, we develop a signal processing framework to explicitly detect such change points in the driver's physiology. Specifically, we apply a multimodal segmentation algorithm to identify change points in measured physiological time series and establish a correspondence to change points in stress-related metrics. We show that pattern changes in physiological signals could be indicative of stress-level changes in drivers and could thus suggest plausible timestamps for stress monitoring, fatigue identification, and just-in-time intervention. We test these premises on three widely used driving behavior analysis benchmarks and perform an extensive ablation study to investigate the robustness of the proposed method. \vspace{-0.25cm}

\section{Biosensing of Driving}
\label{sec:related}
\vspace{-0.05cm}

The automobile industry as well as researchers have developed diverse protocols for sensing and detecting human states while driving. In most cases, experimental data are collected by recruiting subjects to drive, either in the real-world through some pre-defined routes \cite{schneegass2013data}, or through a driving simulator \cite{saeed2017deep}, where the driving settings can be configured. In most cases, both vehicle status and the driver's physiological signals are of interest. Vehicle usage data such as the utilization of pedals and the steering wheel, or meta-parameters like the speed and acceleration of the vehicle can be informative about the state and behavior of the driver \cite{sahayadhas2012detecting}. For the task of arousal estimation, physiological measures are still the most prominent features and multiple signals are typically recorded simultaneously, as single measures could include severe inter-subject variability \cite{hernandez2011call}.

Some of the most widely used sensor measures include electrocardiogram (ECG), electromyogram (EMG), blood pressure, respiration, electrodermal activity (EDA) and skin temperature. Heart rate (HR) and heart rate variability (HRV), derived from the ECG data, are commonly used and have been shown to increase and decrease, respectively, during stress events \cite{kim2018stress, munla2015driver}. HRV can be extracted in either time or frequency domain using various linear or non-linear methods \cite{task1996heart}. Respiration is another indicator for unsafe driving states, such as stress or fatigue \cite{solaz2016drowsiness}. Notably, changes in breathing rate have been correlated to stress episodes \cite{widjaja2013cardiorespiratory}. EMG and temperature measures are more prone to noise and artifacts \cite{chowdhury2018sensor} hence their applicability has been limited. On the other hand, EDA is considered one of the most representative measures due to its connection to the autonomous nervous system activity that causes physiological arousal \cite{stelmack2004psychobiology, affanni2018driver}. EDA captures the exosomatic conductance changes that occur as a result of sweating. Due to the variety of factors that influence the activity of sweat glands e.g., environment temperature, different processing is usually applied. Two signal components that correspond to low-frequency trends and high-frequency oscillations are usually extracted: Skin Conductance Level (SCL) and Skin Conductance Response (SCR). In our study, we follow the literature \cite{stappen2021muse} in considering EDA as a reference (``gold") standard for the drivers' arousal state. \vspace{-0.1cm}

\begin{table*}[t!]
    \footnotesize
    \centering
    \ra{1.2}
    \begin{tabular}{lcccccc}
        \toprule
        \textbf{Dataset} & \textbf{Physiological Signals} & \textbf{\# of Drives} & \textbf{Route Length}  & \textbf{Sample Rate} & \textbf{LP cutoff} & \textbf{GT cutoff} \\ \midrule 
        DriveDB & ECG, Br. Rate, EDA & 24 & $>$ 30 km & 15.5 Hz & 0.05 Hz & 0.01 Hz \\
        HCI Lab Driving & ECG, HR, Skin Temperature, EDA & 10 & About 24 km & 1024 Hz & 0.05 Hz & 0.01 Hz \\
        AffectiveROAD & Br. Rate, Movement, HR, Skin Temperature, EDA & 14 & About 31 km & 1-4 Hz & 0.05 Hz & 0.01 Hz \\ \bottomrule
    \end{tabular}
    \vspace{-0.05cm}
    \caption{Metadata, sampling rates, and downsampling parameters for the datasets used in this study. LP stands for low-pass and GT for ground truth. HR is heart rate, ECG is electrocardiogram, and EDA is electrodermal activity.}
    \label{tab:data_set}
\end{table*}

\section{Change Points In Driving States}
\label{sec:method}
\vspace{-0.1cm}
In this section, we introduce our framework to detect change points from physiological signals recorded during driving. Our pipeline starts with removing high-frequency noises from the physiological time series using low-pass filters. Downsampling is then applied to the filtered time-series data to ensure a common sampling rate for post-analysis. In the final step, we utilize the Gaussian segmentation (GGS) algorithm to partition the time-series data and apply time-series clustering on the derived segments. \vspace{-0.1cm}

\subsection{Data Preprocessing}

For our study, we restrict the scope to HR and respiration measurements, since those are easily available, or can be extracted, in all the aforementioned datasets and have been shown before to be effective indicators of human stress \cite{kim2018stress,widjaja2013cardiorespiratory}. In cases where a dataset does not include respiration measurements, we proceed to extract such information through the ECG-derived Respiration algorithm, as proposed in \cite{van2019heartpy}. From the respiration time-series we then extract the breathing rate using \cite{makowski2021neurokit2}. The physiological measures are originally sampled in various sampling rates, hence we perform downsampling. In specific, we first apply a low-pass, 3rd order Butterworth filter to reduce the high-frequency components and then decimate the filtered signals to a common rate of 0.5 Hz. This way we focus on prominent changes in the signals, while removing possible artifacts. In Table~\ref{tab:data_set} we summarize the sampling parameters used in each of the datasets.

As for the ground truth, we experimented with three different approaches to account for the variability in the provided data. Electrodermal activity (EDA) has been used as a gold standard \cite{stappen2021muse} for affective state estimation, hence we use it as a ground truth measure in all cases. We also evaluate our method on subjective stress ratings whenever available. To enhance the integrity of our work, we additionally consider a reference standard by averaging the time-series of EDA and subjective ratings whenever possible. This strategy has been followed in the literature \cite{stappen2021muse} as a means to ground subjective stress ratings. All ground truth measures are further filtered so as to eliminate any rapid oscillations (Table~\ref{tab:data_set}).

\subsection{Greedy Gaussian Segmentation}

To segment the time-series data we used the greedy Gaussian segmentation (GGS) algorithm, proposed by Hallac et al. \cite{hallac2019greedy}. GGS progressively segments a data stream into parts, whose data points can be described as independent samples of a Gaussian distribution. Given a set of breakpoints $\mathbf{B} = (b_{1}, b_{2}, ..., b_{K})$, the algorithm considers the distribution (mean and co-variance) of the signal changes only at these breakpoints. More specifically, given 2 break points $b_{i}$ and $b_{i+1}$ in $\mathbf{B}$, the GGS estimate the empirical co-variance $S_{i}$ for this segment using the following:
\begin{equation}
    \begin{aligned}
    S^{(i)} = \frac{1}{b_{i} - b_{i-1}} \sum_{t=b_{i-1}}^{b_{i} - 1} (x_{t} - \mu^{(i)})(x_{t} - \mu^{(i)})^{T}
    \end{aligned}
    \label{eqa:segment_covariance}
\end{equation}
where $x_{t}$ is the data sample at $t^{th}$ time point, and $\mu^{(i)}$ and $\Sigma^{(i)}$ are the mean and covariance of the segment, respectively. Using the co-variance computed from the above equation, the GGS tended to estimate $\mathbf{B}$ that maximizes the likelihood shown below:
\begin{equation}
    \begin{aligned}
    -\frac{1}{2} &\sum_{i=1}^{K+1} \Big[ (b_{i} - b_{i-1}) \log\det \Big(S^{(i)} + \frac{\lambda}{b_{i} - b_{i-1}} \mathbf{I}\Big) \\ & - \lambda \textbf{Tr}\Big(S^{(i)} + \frac{\lambda}{b_{i} - b_{i-1}} \mathbf{I}\Big)^{-1} \Big]
    \end{aligned}
    \label{eqa:ggs_likelyhood}
\end{equation}
Here $\lambda$ is a regularization term that sets the importance of the covariance. The problem of searching and deciding over multiple breakpoints is solved with dynamic programming. We intentionally selected GGS as the segmentation algorithm because it can effectively work in multi-modal scenarios by considering multivariate distributions as shown in \cite{feng2020modeling}. Our aim is to detect such breakpoints in the physiological signals and assess their robustness in estimating respective change points of the driver's stress levels.

\subsection{Time-Series Clustering}

\begin{figure*}
    \centering
    \includegraphics[width=0.9\linewidth]{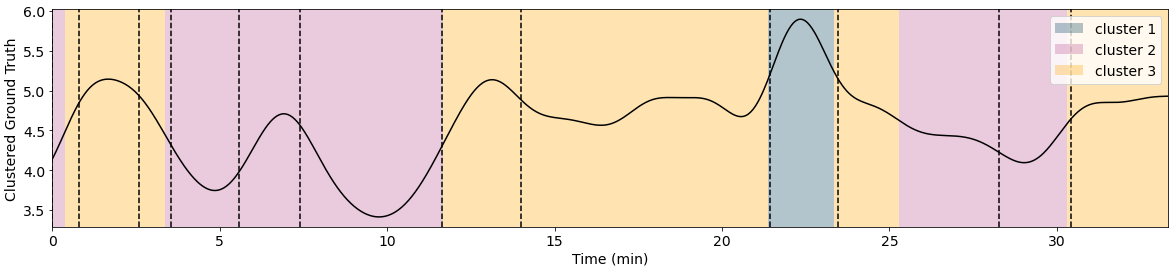}
    \vspace{-0.4cm}
    \caption{Example plot of a clustered EDA ground truth signal from HCI Lab Driving Dataset. The breakpoints denoted by dashed, vertical lines are proposed by the GGS algorithm on the respective heart and breathing rate signals. The covering metric for this case is 0.88.}
    \vspace{-0.3cm}
    \label{fig:example}
\end{figure*}

The primary goal of this study is to detect the most prominent points of change in terms of stress levels but not identify any kind of variation in the data. To quantify this notion, we applied temporal clustering to the ground truth time-series, based on the breakpoints proposed by the GGS algorithm. After running GGS, we regarded each segment as an independent time-series sample and performed simple time-series k-means to cluster the segments into $k$ sets. Subsequently, we use the clustering results to discard all breakpoints lying between segments of the same cluster. A visualization of the resulting representation for a sample EDA signal can be seen in Figure~\ref{fig:example}.

\section{Experimental Setup}

\subsection{Dataset}

Data sources for this kind of study are limited, both because of challenges in the data collection process and privacy issues regarding the collected measures. To the best of our knowledge, there is still no publicly available dataset that contains multiple physiological signals, alongside data with respect to driving (e.g., speed, pedal utilization), and also fine-grained stress annotations. For our study, we used three public datasets that aligned most with our goals.

\textbf{DriveDB} dataset \cite{healey2005detecting}, available through PhysioNet \cite{moody2001physionet}, is collected from 24 drives with a minimum of 50-min duration over a 32.2 km driving session. The dataset provides detailed physiological measurements of ECG, EMG, EDA, and breathing rate from the driver. Each driver follows a set of pre-selected routes including both highway and city driving conditions. The objective of the study is to detect stress levels from physiological responses during driving, however the stress ratings are not released within the dataset. The raw ECG data has a sampling rate of 496 Hz, and the HR, EDA and respiration have a sampling rate of 15.5 Hz.

\textbf{HCI Lab Driving} database \cite{schneegass2013data} was created to assess driver workload in a real-world driving setting. The data are collected from 10 participants, 7 male, and 3 female. Each participant drove a selected route covering around 23.6 km with instructions given by the researcher. Each driving route has five different road types: highway, freeway, tunnel, 30 km/h zone, and 50 km/h zone. The corpus contains GPS, physiological data, and acceleration data under real-world driving setups. The recorded physiological data contains measurements of ECG, heart rate, skin temperature, and EDA. In particular, heart rate, body temperature, and EDA data were sampled at 128Hz, while the ECG data was collected at 1024 Hz. Here, stress annotations are also provided through a video rating session, performed by the subject after the driving session and range between 0 (no workload) to 128 (maximum workload), recorded at 29 FPS on average. All data streams were extrapolated to 1024 Hz.

\textbf{AffectiveROAD} dataset \cite{haouij2018affectiveroad} is a multimodal driving database of physiological data and ambient environment data collected from 10 participants (five male and five female). The final set contains 14 drives, and each drive followed a set of pre-selected routes with a total length of about 31 km containing both city conditions and highway conditions. The physiological signals were collected using two Empatica E4 wristband sensors and a chest-wrap wearable sensor called Zephyr Bioharness 3. These sensors offer non-obtrusive measures of EDA, skin temperature, HR, triaxial acceleration, and breathing rate at sampling rates of 1-4 Hz. \vspace{-0.2cm}

\begin{table*}[t!]
    \footnotesize
    \centering
    \ra{1.1}
    \begin{tabular}{lccccccccc} \toprule
    \multirow{2}{*}{\textbf{Dataset}} & \multicolumn{3}{c}{\textbf{EDA Ground Truth}} & \multicolumn{3}{c}{\textbf{SR Ground Truth}} & \multicolumn{3}{c}{\textbf{EDA$+$SR Ground Truth}}
    \\ & \textbf{HR} & \textbf{BR} & \textbf{Multimodal}
     & \textbf{HR} & \textbf{BR} & \textbf{Multimodal} & \textbf{HR} & \textbf{BR} & \textbf{Multimodal} \\
    \cmidrule(lr){1-1} \cmidrule(lr){2-4} \cmidrule(lr){5-7} \cmidrule(lr){8-10}
    DriveDB & \textbf{0.80}$\pm$0.10 & 0.78$\pm$0.09 & 0.76$\pm$0.12 & -- & -- & -- & -- & -- & --\\
    HCI Lab Driving & \textbf{0.82}$\pm$0.09 & -- & -- & 0.78$\pm$0.09 & -- & -- & 0.75$\pm$0.14 & -- & -- \\
    AffectiveROAD & 0.67$\pm$0.10 & 0.70$\pm$0.09 & 0.42$\pm$0.16 & 0.81$\pm$0.15 & 0.82$\pm$0.10 & 0.61$\pm$0.21 & \textbf{0.94}$\pm$0.04 & 0.91$\pm$0.06 & 0.89$\pm$0.05\\
    \bottomrule
    \end{tabular}
    \vspace{-0.15cm}
    \caption{Average covering metric \cite{arbelaez2010contour} results for each dataset and signal (heart rate -- HR, breathing rate -- BR, HR+BR -- Multimodal), tested on Electrodermal Activity (EDA), Stress Rating (SR), and fused (SR + EDA) ground truths. $\pm$ denotes 1 standard deviation.}
    \vspace{-0.4cm}
    \label{tab:eda_results}
\end{table*}

\subsection{Evaluation Protocol}

There are multiple metrics to evaluate a point detection problem. In our study that focuses on time-series clustering, we use the \textit{covering} metric, as described in \cite{arbelaez2010contour}. Let $S = \{b_1, \dots, b_n\}$ denote the set of locations of $n$ breakpoints provided in a ground truth time-series of length T. $S$ implies a partition $G$ of the time-series into segments, where $A_j$ is the segment from $b_{j-1}$ to $b_j - 1$ for $j = 2, \dots, n$. Similarly, the GGS algorithm proposes a set of breakpoints $S^\prime$ to partition the ground truth. The covering metric utilizes the Jaccard index $J\left(A,A^\prime\right)$ between a ground truth and a proposed set, $A$, $A^\prime$. Then, the covering metric of a partition $G$ by a partition $G^\prime$ is defined as the normalized sum of the similarity scores $J(A, A^\prime)$ over all ground-truth segments, where for each segment we consider the set that provides the maximum index out of all intervals that can be derived from the proposed breakpoints:
\begin{equation}
    Cover\left(G, G^{\prime}\right)=\frac{1}{T} \sum_{A \in G}|A| \max _{A^{\prime} \in G^{\prime}} J\left(A, A^{\prime}\right)
\end{equation}
As a simple baseline for our experiments, we consider the covering metric of the entire time-series being the only available proposal.\vspace{-0.1cm}

\subsection{Change Point Detection}

As described in section \ref{sec:method}, we first apply a low-pass, 3rd-order Butterworth filter to reduce the high-frequency components with a cutting frequency of 0.05Hz (Table~\ref{tab:data_set}). We then re-sample each time-series data to a common rate of 0.5Hz. Ground truth time-series are further downsampled at a cutting frequency of 0.01Hz to preserve only prominent arousal changes. The GGS algorithm is then used to extract a set of break-points with $\lambda=15$, whereas the maximum number of breakpoints is determined as a function of the time-series length, at a rate of 15 break-points per hour. \vspace{-0.1cm}

\begin{figure}
    \centering
    \includegraphics[scale=0.365]{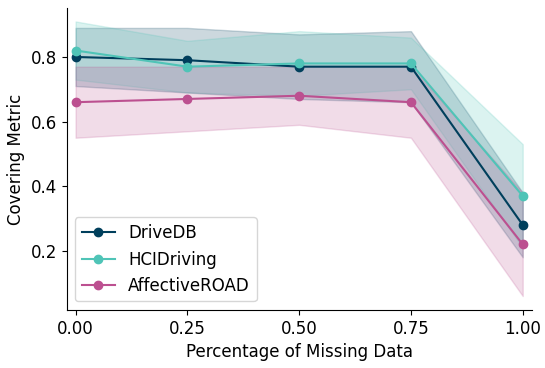}
    \vspace{-0.3cm}
    \caption{Ablation study in which we mask random sub-sequences from the raw signals, up to 1\% of continuous data points. For fair comparison, we solely include HR proposals and EDA ground truth.}
    \label{fig:missing}
\end{figure}

\begin{table}[t!]
    \footnotesize
    \centering
    \ra{1.1}
    \begin{tabular}{lccc}
        \toprule
        \textbf{Dataset} & $\lambda = 1$  & $\lambda = 5$ & $\lambda = 15$\\ \midrule 
        DriveDB & 0.80 $\pm$ 0.11 & \textbf{0.81 $\pm$ 0.09} & 0.80 $\pm$ 0.10 \\
        HCI Lab Driving & 0.76 $\pm$ 0.10 & 0.76 $\pm$ 0.10 & \textbf{0.82 $\pm$ 0.09} \\
        AffectiveROAD & 0.65 $\pm$ 0.10 & 0.66 $\pm$ 0.12 & \textbf{0.67 $\pm$ 0.10} \\ \bottomrule
    \end{tabular}
    \vspace{-0.1cm}
    \caption{Ablation on the $\lambda$ parameter of the segmentation algorithm.}
    \vspace{-0.4cm}
    \label{tab:lambda}
\end{table}

\section{Experimental Results}

The experimental results of monitoring the heart and breathing rate measurements for the 3 available datasets are summarized in Table~\ref{tab:eda_results}. We observe that the covering metric is satisfactorily high in nearly all cases, indicating that changes in physiological measures can indeed point to prominent changes in the arousal state. Considering EDA as ground truth, we report a covering metric above 70\% for both DriveDB and HCI Lab Driving. In the first case, heart rate shows slightly better performance, however if we consider both HR and breathing rate, performance drops and variance increases. This behavior is consistent in all experiments and indicates that the relative contribution of each measure varies between subjects.

Regarding AffectiveROAD, the monitored signals do not align as well as in the previous cases, limiting the obtained covering metric to 67\% for the heart rate and 70\% for the breathing rate. In contrast, when considering the explicit stress annotations as ground truth, both signals are robust in locating state changes, at a comparable accuracy. In the third version of our experiments, where we considered a fused gold standard of reported ratings and EDA, AffectiveROAD gets a substantial boost of 27\% percentage points, compared to EDA-only performance and 13\% compared to the rating-only performance, indicating that fusion of that kind efficiently grounds arousal estimations in AffeciveROAD. We underscore that the performance of the multimodal model doubles its score compared to those from the EDA-only ones.

Our base experiment includes a $\lambda$ parameter of 15 for the segmentation. Ablative experiments, shown in Table~\ref{tab:lambda}, show that higher values of this parameter generally provide better outcomes. As presented in equation~\ref{eqa:ggs_likelyhood}, lower values of $\lambda$ would reduce regularization and thus result in a larger number of proposed breakpoints that are sensitive to noisy elements of the signals. By using higher values of $\lambda$ we focus instead on more prominent changes that are typically shared between the monitored signal and the ground truth. \vspace{-0.1cm}

\subsection{Missing Data Setting}

We further investigate the robustness of our algorithm by simulating missing data points in the monitored signals. Missing data occurs frequently in physiological data collection \cite{dong2019improved}, especially in dynamic settings like driving, where the sensors might disconnect or produce motion artifacts. Our simulation includes masking the raw monitored signals with random patches of at most 1\% the length of the respective signal. The missing points are then imputed with the last available signal value. We present our results in Figure~\ref{fig:missing} for all 3 datasets. We observe that both the algorithm performance and its variance remain stable even when 50\% of the data are missing. Performance gradually drops under 40\% when all data are masked, giving us a rough baseline for all derived scores. It should be noted, though, that robustness in certain cases results from pre-configured upsampling of the raw signals within the dataset. \vspace{-0.15cm}

\subsection{Number of Clusters}

\begin{figure}
    \centering
    \includegraphics[scale=0.373]{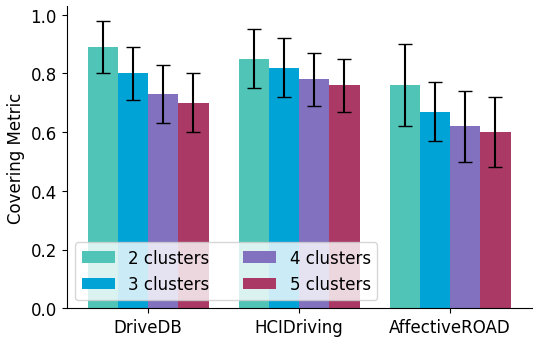}
    \vspace{-0.3cm}
    \caption{Ablation of varying the number of ground truth clusters. EDA is considered the ground truth and HR is the monitored signal.}
    \label{fig:clusters}
    \vspace{-0.25cm}
\end{figure}

We additionally experimented with the number of assigned clusters in the ground truth signal. This is a crucial parameter in determining which change points will be selected as prominent. While our initial configuration accounted for the distribution of low -- medium -- high arousal, other configurations are possible, hence we list their performance in Figure~\ref{fig:clusters}. As expected, performance gradually drops as the number of clusters increases, but remains satisfactorily high, above 70\% in most cases, even in the scenario of 5 arousal levels.

\section{Conclusion}
\label{sec:conc}
\vspace{-0.1cm}
In this paper we examined to what extent multimodal physiological signals like HR and breathing rate could be indicative of prominent changes in physiological arousal during driving. To that end, we developed a framework to filter the signals, identify trends of interest and determine breakpoints using time-series segmentation. The proposed algorithm was applied against both reference EDA and self-reported stress ratings, showing that breakpoint proposals from the monitored signals highly align with points of arousal changes. We hope these results support further research into robust localization of unsafe driving states and enhance just-in-time intervention systems. Future work will address current limitations including the number of false-positive proposals and the efficacy of multimodal monitoring.

\newpage
\bibliographystyle{IEEEbib}
\bibliography{refs}

\end{document}